\documentclass[a4paper,11pt]{article}
\usepackage[margin=2.5cm]{geometry}
\usepackage{lineno}
\usepackage{amsfonts,amsmath,amssymb}
\usepackage[titletoc,toc,title]{appendix}
\usepackage{authblk}
\usepackage{braket}
\usepackage{cancel}
\usepackage[font={footnotesize,it}]{caption}
\usepackage{cite}
\usepackage{comment}
\usepackage{enumitem}
\usepackage{epsfig}
\usepackage{etoolbox}
\usepackage{float}
\usepackage{graphicx}
\usepackage{epstopdf}
\usepackage{physics}
\usepackage{ragged2e}
\usepackage{setspace}
\usepackage{soul}
\usepackage[raggedright]{titlesec}
\usepackage{xcolor}
\usepackage{hyperref}
\usepackage{cleveref}

\newcommand*\linenomathpatch[1]{%
  \cspreto{#1}{\linenomath}%
  \cspreto{#1*}{\linenomath}%
  \csappto{end#1}{\endlinenomath}%
  \csappto{end#1*}{\endlinenomath}%
}
\linenomathpatch{equation}
\linenomathpatch{align}
\hypersetup{colorlinks=true,citecolor=red,linkcolor=blue,urlcolor=blue,linktocpage=true}
{\makeatletter\g@addto@macro\bfseries{\boldmath}\makeatother}
\numberwithin{equation}{section}
\crefname{subsection}{subsection}{subsections}
\def\cft#1{CFT$_{#1}$}
\def\ads#1{AdS$_{#1}$}

\begin{document}
\title{Holographic entanglement negativity for disjoint subsystems in AdS$_{d+1}$/CFT$_d$}
\author[1,2]{Jaydeep Kumar Basak\thanks{E-mail: \texttt{jaydeep@mail.nsysu.edu.tw}}}
\author[3,4]{Himanshu Parihar\thanks{E-mail: \texttt{himansp@phys.ncts.ntu.edu.tw}}}
\author[5]{Boudhayan Paul\thanks{E-mail: \texttt{paul@bimsa.cn}}}
\author[6]{Gautam Sengupta\thanks{E-mail: \texttt{sengupta@iitk.ac.in}}}
\affil[1]{Department of Physics, National Sun Yat-Sen University, Kaohsiung 80424, Taiwan
\medskip}
\affil[2]{Center for Theoretical and Computational Physics, Kaohsiung 80424, Taiwan
\medskip}
\affil[3]{Center of Theory and Computation, National Tsing-Hua University, Hsinchu 30013, Taiwan
\medskip}
\affil[4]{Physics Division, National Center for Theoretical Sciences, Taipei 10617, Taiwan
\medskip}
\affil[5]{Beijing Institute of Mathematical Sciences and Applications, Beijing 101408, China
\medskip}
\affil[6]{Department of Physics, Indian Institute of Technology, Kanpur 208016, India}
\date{}
\maketitle
\begin{abstract}
We propose a construction to compute the holographic entanglement negativity for bipartite mixed state configurations of two disjoint subsystems in higher dimensional conformal field theories (CFT$_d$s) dual to bulk AdS$_{d+1}$ geometries. Our proposal follows from the corresponding AdS$_3$/CFT$_2$ scenario and involves an algebraic sum of the areas of bulk RT surfaces for certain combinations of subsystems. Utilizing our construction we compute the holographic entanglement negativity at the leading order through a perturbative expansion, for such bipartite mixed states of two disjoint subsystems with long rectangular strip geometries in CFT$_d$s dual to bulk pure AdS$_{d+1}$ geometries and AdS$_{d+1}$-Schwarzschild black holes.
\end{abstract}
\clearpage
\begingroup
\RaggedRight
\tableofcontents
\endgroup
\clearpage

\section{Introduction}\label{sn_intro}

In recent years quantum entanglement has emerged as a central issue in a diverse range of phenomena from condensed matter physics to quantum gravity and black holes. For bipartite pure states, quantum entanglement is characterized through the entanglement entropy which is the von Neumann entropy of the reduced density matrix for the subsystem in question. For two dimensional conformal field theories (\cft{2}s) the entanglement entropy for bipartite states may be explicitly computed through the replica technique as described in \cite{Calabrese:2004eu,Calabrese:2009ez,Calabrese:2009qy,Calabrese:2010he}. For bipartite mixed states however the entanglement entropy receives contributions from irrelevant correlations which renders it invalid as an entanglement measure for such states. Several entanglement and correlation measures have been introduced in quantum information theory for characterizing mixed state entanglement although most of these are not easily computable. A computable measure termed as entanglement negativity (logarithmic negativity) for characterization of mixed state entanglement was introduced in a classic work by Vidal and Werner \cite{Vidal:2002zz}. This was defined as the logarithm of the trace norm for the partially transposed density matrix for a bipartite system with respect to one of the subsystems and characterized an upper bound to the distillable entanglement of the mixed state. Although shown to be non convex by Plenio \cite{Plenio:2005cwa} it was found to be an entanglement monotone and hence constituted a measure for mixed state entanglement. Interestingly as described in \cite{Calabrese:2012ew,Calabrese:2012nk,Calabrese:2014yza} the entanglement negativity for bipartite pure and mixed states in \cft{2}s could also be explicitly computed through a replica technique.

In the above context Ryu and Takayanagi (RT) \cite{Ryu:2006bv,Ryu:2006ef} proposed a holographic conjecture to obtain the entanglement entropy for bipartite states in $d$ dimensional \cft{}s (\cft{d}s) dual to bulk \ads{d+1} geometries in the context of the \ads{d+1}/\cft{d} correspondence. The holographic entanglement entropy was shown to be proportional to the area of a codimension two bulk static minimal surface (RT surface) homologous to the subsystem in question. Their conjecture was utilized to compute the entanglement entropy for various bipartite states in holographic \cft{}s \cite{Nishioka:2009un,Takayanagi:2012kg,Cadoni:2010kla,Fischler:2012ca,Fischler:2012uv,Blanco:2013joa,Chaturvedi:2016kbk, Nishioka:2018khk}. Subsequently a covariant generalization of the RT conjecture was established by Hubeny, Rangamani and Takayanagi (HRT) in \cite{Hubeny:2007xt}. Explicit proofs of the RT and the HRT conjectures were subsequently established in a series of works \cite{Faulkner:2013yia,Lewkowycz:2013nqa,Dong:2016hjy}.

The developments described above naturally led to the significant issue of a holographic description of entanglement negativity for bipartite states in a dual \cft{d} which was critical to understand mixed state entanglement in AdS/CFT. In this context a holographic computation for the entanglement negativity of bipartite pure vacuum states in \cft{d}s was described in \cite{Rangamani:2014ywa} for a generic \ads{d+1}/\cft{d} scenario. However a holographic construction for general bipartite states (including mixed states) in dual \cft{d}s remained an outstanding issue. Interestingly a holographic conjecture for the entanglement negativity of bipartite states in \cft{2}s and its covariant extension were proposed in \cite{Chaturvedi:2016rcn,Chaturvedi:2017znc} for the \ads{3}/\cft{2} scenario which reproduced the corresponding field theory replica technique results in the large central charge limit. This construction involved an algebraic sum of the areas of bulk RT surfaces (lengths of bulk geodesics in this case) for certain combinations of the intervals for the bipartite state in question. A rigorous large central charge analysis for the entanglement negativity in dual \cft{2}s utilizing the monodromy technique was subsequently developed in \cite{Malvimat:2017yaj} which provided a strong consistency check for this conjecture. This holographic proposal and its covariant extension were further developed in \cft{2}s dual to bulk \ads{3} geometries for two adjacent intervals in \cite{Jain:2017aqk,Jain:2017uhe}, and for two disjoint intervals in proximity\footnote{The proximity regime is realized when the separation between the two disjoint intervals becomes much smaller than the sizes of the intervals \cite{Malvimat:2018txq,Malvimat:2018ood}.} in \cite{Malvimat:2018txq,Malvimat:2018ood}.

Higher dimensional generalizations of the above holographic entanglement negativity conjectures for the \ads{d+1}/\cft{d} framework were described in another series of articles in \cite{Chaturvedi:2016rft,Jain:2017xsu,Jain:2018bai}. These involved corresponding algebraic sums of the areas of bulk RT surfaces for certain combinations of the subsystems for the bipartite state in question. These conjectures were applied to compute the holographic entanglement negativity for bipartite states involving single and adjacent subsystems with long\footnote{\label{fn_long}The term long rectangular strip means that one spatial dimension of the rectangular subsystem is much smaller than all other spatial dimensions.} rectangular strip geometries in \cft{d}s dual to bulk pure \ads{d+1} geometries and \ads{d+1}-Schwarzschild and \ads{d+1}-Reissner-Nordstr{\"o}m black holes respectively to leading orders in a perturbative expansion. Interestingly these results reproduced certain interesting features such as the elimination of thermal terms in the holographic entanglement negativity for corresponding lower dimensional \ads{3}/\cft{2} scenario in the literature \cite{Chaturvedi:2016rcn,Chaturvedi:2016rft}. This conformed to the characterization of the entanglement negativity as an upper bound to the distillable entanglement in quantum information theory and constituted crucial consistency checks.

In this article we extend the higher dimensional holographic construction for the entanglement negativity to disjoint subsystems. This is expected to provide further substantiation to the holographic characterization for the entanglement negativity in dual \cft{d}s and the development of a proof from a bulk gravitational path integral perspective.\footnote{A plausible proof of our holographic conjecture for spherical entangling surfaces was recently described in the appendix of \cite{KumarBasak:2020ams} following the work in \cite{Dong:2021clv} on replica symmetry breaking saddles for the bulk gravitational path integral although it still remains an open issue for generic subsystem geometries.} To this end we extend the holographic entanglement negativity conjecture for mixed state configurations of disjoint intervals (in proximity) in the \ads{3}/\cft{2} scenario described in \cite{Malvimat:2018txq,Malvimat:2018ood} to a higher dimensional \ads{d+1}/\cft{d} framework in terms of a similar algebraic sum of the areas of bulk RT surfaces for corresponding combinations of the subsystems.\footnote{The characterization of the proximity regime for higher dimensional scenarios is described in \cref{sn_hen_hd_zero} in \cref{proximity} and the discussion preceding it.}

In this connection we utilize the above higher dimensional construction to obtain the holographic entanglement negativity of bipartite states of two disjoint subsystems (in proximity) with long rectangular strip geometries in \cft{d}s dual to bulk pure \ads{d+1} space time and the \ads{d+1}-Schwarzschild black holes respectively to leading orders in a perturbation theory.\footnote{Note that in \cite{Erdmenger:2017pfh} an exact evaluation of this has been provided through the Meijer G-functions however we adopt a perturbative approach in this article which provides the necessary leading order behavior for the area functional.} Interestingly the holographic entanglement negativity at leading orders is cutoff independent and depends only on the area of the entangling surfaces (no volume  dependent thermal terms) as expected from the corresponding lower dimensional \ads{3}/\cft{2} results \cite{Malvimat:2018txq,Malvimat:2018ood}. We also observe that in the limit where the subsystems are adjacent, our results exactly reproduce earlier results described in \cite{Jain:2017xsu} providing a consistency check for the holographic construction.

Note that the authors in \cite{Hung:2011nu,Hung:2014npa} have studied the replica technique for the entanglement entropy of the vacuum state in flat higher dimensional conformal field theories involving twist operators for spherical entangling surfaces and computed their conformal dimensions. For specific bulk gravitational theories they demonstrated the consistency of their computations through comparison with the corresponding holographic results. However the extension of such higher dimensional twist fields for entangling surfaces of arbitrary geometries is a technically challenging outstanding issue. Moreover the computations in \cite{Hung:2011nu,Hung:2014npa} are only valid for the pure vacuum state of flat higher dimensional CFTs. The generalization of their computations to mixed states for the entanglement negativity and for entangling surfaces with non spherical geometries is hence still an open question for future investigations.

We should mention here that an alternate holographic entanglement negativity conjecture involving the entanglement wedge cross section (EWCS) backreacted by a cosmic brane for the replicated bulk conical defect geometry has been developed in \cite{Shapourian:2018lsz,Kudler-Flam:2018qjo,Kudler-Flam:2019wtv,Kusuki:2019zsp} and further elucidated in \cite{KumarBasak:2020eia}. For the \ads{3}/\cft{2} scenario this holographic proposal reproduced the entanglement negativity for bipartite states in \cft{2}s dual to bulk \ads{3} geometries in the large central charge limit modulo certain constants involving the Markov gap \cite{Hayden:2021gno}. For spherical entangling surfaces, an overall dimension dependent constant backreaction factor $\mathcal{X}_d$ for the bulk cosmic brane may be determined from the entanglement negativity of the corresponding vacuum state in \cft{d}s dual to a bulk pure \ads{d+1} geometries \cite{Casini:2011kv,Hung:2011nu,Rangamani:2014ywa}.\footnote{Note that $\mathcal{X}_d=(1/2)x_d^{d-2}(1+x_d^2)-1$, where $x_d=(2/d)\left(1+\sqrt{1-d/2+d^2/4}\right)$. For details see \cite{Casini:2011kv,Hung:2011nu,Rangamani:2014ywa,Kudler-Flam:2018qjo,Kusuki:2019zsp}.} Keeping these developments in perspective, we state here that our higher dimensional results reported in this article will also involve backreaction effects due to the bulk cosmic brane for the replicated conical defect geometry. However the determination of these backreaction effects for the long rectangular subsystem geometries discussed in this article remains a non trivial open issue.

This article is organized as follows. In \cref{sn_hen_3d} we briefly review the computation of holographic entanglement negativity for bipartite mixed state configuration of two disjoint intervals in the context of the \ads{3}/\cft{2} framework described in \cite{Malvimat:2018txq}. Subsequently in \cref{sn_hen_hd} we propose a holographic entanglement negativity conjecture for mixed states of two disjoint subsystems in the \ads{d+1}/\cft{d} scenario, and describe its application to the zero temperature mixed state of two disjoint subsystems with long rectangular strip geometries in a \cft{d} dual to a bulk pure \ads{d+1} geometry in \cref{sn_hen_hd_zero}. Following this in \cref{sn_hen_hd_finite} we utilize our proposal to compute the holographic entanglement negativity for such mixed states at finite temperatures in \cft{d}s dual to bulk \ads{d+1}-Schwarzschild black holes. Finally we summarize our results in \cref{sn_sum} and present our conclusions.

\section{Holographic entanglement negativity for \ads{3}/\cft{2}}\label{sn_hen_3d}

In this section we briefly review the computation of the entanglement negativity for bipartite states of two disjoint intervals in a \cft{2} through a replica technique in the large central charge limit \cite{Calabrese:2012ew,Calabrese:2012nk,Calabrese:2014yza}. Subsequently we describe the computation of the holographic entanglement negativity for such configurations in \cft{2}s dual to bulk pure \ads{3} and the BTZ black hole geometries \cite{Malvimat:2018txq}.

\subsection{Entanglement negativity in a \cft{2}}

\begin{figure}
\begin{center}
\includegraphics[scale=0.9]{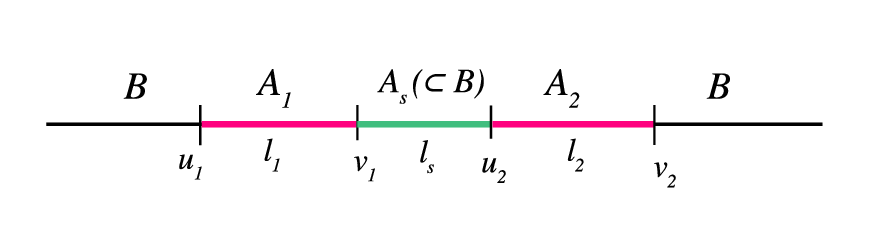}
\caption{Two disjoint intervals $A_1$ and $A_2$ with rest of the system $B$ in a \cft{2}.}
\label{fig_en_dj}
\end{center}
\end{figure}

We begin by considering a tripartition of a system in a pure state described by a \cft{2} into the spatial intervals $A_1$, $A_2$ and $B$ with $A=A_1\cup A_2=[u_1,v_1]\cup[u_2,v_2]$, and $B=A^c$ represents the rest of the system. The two intervals $A_1$ and $A_2$ are separated by the interval $ A_s\subset B$ as shown in \cref{fig_en_dj}.

The reduced density matrix for the subsystem $A$ which is in a mixed state is defined by tracing over the degrees of freedom of $B$ as $\rho_A=\mathrm{Tr}_{B}(\rho_{AB})$, where $\rho_{AB}$ is the density matrix for the full system $A\cup B\equiv AB$. The partial transpose $\rho_A^{T_{A_2}}$ of the reduced density matrix $\rho_A$ with respect to the subsystem $A_2$ is defined by
\begin{equation}
\label{pt_def}
\mel{e^{(A_1)}_ie^{(A_2)}_j}{\rho_A^{T_{A_2}}}{e^{(A_1)}_ke^{(A_2)}_l}=\mel{e^{(A_1)}_ie^{(A_2)}_l}{\rho_A}{e^{(A_1)}_ke^{(A_2)}_j},
\end{equation}
where the orthonormal bases characterizing the Hilbert spaces $\mathcal{H}_{A_1}$ and $\mathcal{H}_{A_2}$ are denoted by ${|e^{(A_1)}_i\rangle}$ and $|e^{(A_2)}_j\rangle$ respectively.

The entanglement negativity $\mathcal{E}$ between the disjoint intervals $A_1$ and $A_2$ (also referred to as the entanglement negativity of the subsystem $A$) may then defined as the logarithm of the trace norm of the partially transposed reduced density matrix as\footnote{The trace norm of any arbitrary hermitian matrix $M$ is defined as $|M|=\text{Tr}\sqrt{MM^\dagger}$.}
\begin{equation}\label{negativity}
\mathcal{E} = \ln\mathrm{Tr}\left|\rho_A^{T_{A_2}}\right|.
\end{equation}
The entanglement negativity may now be obtained through a replica technique as discussed in \cite{Calabrese:2012ew,Calabrese:2012nk} to determine $\mathrm{Tr}(\rho_A^{T_{A_2}})^{n_e}$ and the replica limit is given as the analytic continuation of even sequences of $n=n_e$ to $n_e\to 1$. This leads us to the following expression for the entanglement negativity of the subsystem $A$ as
\begin{equation}\label{en}
\mathcal{E}=\lim_{n_e\to 1}\ln\mathrm{Tr}\left(\rho_A^{T_{A_2}}\right)^{n_e}.
\end{equation}
For the configuration of the two disjoint intervals as shown in \cref{fig_en_dj}, the quantity $\mathrm{Tr}(\rho_A^{T_2})^{n_e}$ in \cref{en} is expressed as a four point correlator of the twist fields on the complex plane $\mathbb{C}$ in the following way
\begin{equation}\label{four-point}
\mathrm{Tr}(\rho_A^{T_{A_2}})^{n_e}=\langle\mathcal{T}_{n_e}(u_1)\overline{\mathcal{T}}_{n_e}(v_1)
\overline{\mathcal{T}}_{n_e}(u_2)\mathcal{T}_{n_e}(v_2)\rangle_{\mathbb{C}}.
\end{equation}

In the replica limit $n_e\to 1$, the four point function of the twist fields described above is expressed in \cite{Calabrese:2012nk} as
\begin{equation}\label{four-point_Replica}
\lim_{n_e\to 1}
\left\langle{\cal T}_{n_e}(u_1)
\overline{{\cal T}}_{n_e}(v_1)
\overline{{\cal T}}_{n_e}(u_2)
{\cal T}_{n_e}(v_2)\right\rangle_{\mathbb{C}}
= {\cal G}(x).
\end{equation}
The arbitrary function ${\cal G}(x)$ of the cross ratio $x=(v_1-u_1)(v_2-u_2)/[(u_2-u_1)(v_2-v_1)]$ in \cref{four-point_Replica} is non universal and depends on the full operator content of the \cft{2}. We briefly describe a procedure to obtain an expression for the above four point twist correlator in the large central charge limit through the monodromy technique \cite{Fitzpatrick:2014vua,Hartman:2013mia,Kulaxizi:2014nma,Malvimat:2018txq}.

\subsection{Entanglement negativity at large central charge}

The four point function in \cref{four-point_Replica} in the large central charge limit when the two disjoint intervals are in proximity ($x\approx 1$) may be expressed as \cite{Hartman:2013mia,Kulaxizi:2014nma,Malvimat:2017yaj}
\begin{equation}\label{four-point-limit}
\lim_{n_e\to 1}\left\langle{\cal T}_{n_e}(z_1){\overline{\cal T}}_{n_e}(z_2)
{\overline{\cal T}}_{n_e}(z_3){\cal T}_{n_e}(z_4)\right\rangle_\mathbb{C}=\left(1-x\right)^{2\hat{h}},
\end{equation}
where we have identified $u_1\equiv z_1,v_1\equiv z_2,u_2\equiv z_3,v_2\equiv z_4$ and the cross ratio $ x $ is given by $(z_{12}z_{34})/(z_{13}z_{24})$ with $z_{ij}\equiv z_i-z_j$, and $\hat{h}$ is the conformal dimension of the operator with the dominant contribution in the corresponding conformal block expansion. The entanglement negativity for the bipartite mixed state configuration of disjoint intervals in proximity\footnote{Note that when the two disjoint intervals are far away ($x\approx 0$), the negativity vanishes non perturbatively \cite{Calabrese:2012nk}.} may then be obtained from \cref{en,four-point-limit} as \cite{Malvimat:2018txq}
\begin{equation}\label{negativity-disjoint}
\mathcal{E}=\frac{c}{4}\ln\left(\frac{|z_{13}|\;|z_{24}|}{|z_{14}|\;|z_{23}|}\right).
\end{equation}
The entanglement negativity for the zero temperature mixed state of the two disjoint intervals in proximity is then determined from \cref{negativity-disjoint} by substituting the lengths of the respective intervals as follows
\begin{equation}\label{en_disj_vac}
\mathcal{E}=\frac{c}{4}\ln\left[\frac{(l_1+l_s)(l_2+l_s)}{l_s(l_1+l_2+l_s)}\right].
\end{equation}
Interestingly note however that the result described in the above equation is cutoff independent in contrast to that for the mixed state of adjacent intervals \cite{Calabrese:2012nk}.

The entanglement negativity for the corresponding mixed state of disjoint intervals at a finite temperature $T$ at large $c$ may be obtained as above through the conformal map $z\to w=(\beta/2\pi)\ln z$ from the complex plane to the cylinder where the Euclidean time direction has now been compactified to a circle with circumference $\beta\equiv 1/T$ \cite{Calabrese:2014yza}. The four point function of the twist fields in \cref{four-point} transforms under the conformal map described above as follows
\begin{align}
\left\langle{\cal T}_{n_e}(w_1)\overline{{\cal T}}_{n_e}(w_2)
\overline{{\cal T}}_{n_e}(w_3){\cal T}_{n_e} (w_4)\right\rangle_{cyl}
=\prod_{i=1}^{4}&\left[\left(\frac{dw(z)}{dz}\right)^{-\Delta_i}\right]_{z=z_i}\nonumber\\
&\times\left\langle{\cal T}_{n_e}(z_1)\overline{{\cal T}}_{n_e}(z_2)
\overline{{\cal T}}_{n_e}(z_3){\cal T}_{n_e}(z_4)\right\rangle_\mathbb{C},\label{finite-trans}
\end{align}
where $\Delta_i$ are the scaling dimensions of the twist fields at the locations $w=w_i$.

The entanglement negativity for the mixed state configuration of disjoint intervals in proximity at a finite temperature in the limit of large central charge $c$ is then obtained from \cref{four-point-limit} upon utilizing \cref{en,four-point,finite-trans}, as follows
\begin{equation}\label{cft-finite-temp}
\mathcal{E}=\frac{c}{4}\ln
\left[\frac{\sinh\{\pi(l_1+l_s)/\beta\}\sinh\{\pi(l_2+l_s)/\beta\}}
{\sinh\{\pi l_s/\beta\}\sinh\{\pi(l_1+l_2+l_s)/\beta\}}\right].
\end{equation}
The above result is also cutoff independent as earlier. Interestingly for the mixed state configuration of two disjoint intervals in proximity it could be numerically established that the entanglement negativity exhibits a phase transition \cite{Kulaxizi:2014nma,Dong:2018esp}.

\subsection{Holographic entanglement negativity for two disjoint intervals in a \cft{2}}

Following the derivation of the entanglement negativity in the large central charge limit for mixed states of two disjoint intervals in proximity, we now outline the corresponding holographic construction as described in \cite{Malvimat:2018txq}. The two point function of the twist fields in a holographic \cft{2} can be expressed as
\begin{equation}\label{H-2-point}
\left\langle{\cal T}_{n_e}(z_i)\overline{{\cal T}}_{n_e}(z_j)\right\rangle_\mathbb{C}
\sim\left|z_{ij}\right|^{-2\Delta_{{\cal T}_{n_e}}}.
\end{equation}
From the \ads{3}/\cft{2} lexicon, the two point function of the twist fields in \cref{H-2-point} (in the geodesic approximation) is described in terms of of the length ${\cal L}_{ij}$ of the bulk space like geodesic which is homologous to the interval in question as \cite{Ryu:2006ef}
\begin{equation}\label{H-2-bulk}
\left\langle{\cal T}_{n_e}(z_i)\overline{{\cal T}}_{n_e}(z_j)\right\rangle_\mathbb{C}
\sim\exp\left(-\Delta_{{\cal T}_{n_e}}{\cal L}_{ij}/R\right),
\end{equation}
where $R$ is the \ads{3} length scale.\\
Using \cref{H-2-point,H-2-bulk}, the four point twist correlator in \cref{four-point-limit} may be written as
\begin{equation}\label{HenConjecture}
\lim_{n_e\to 1}
\left\langle{\cal T}_{n_e}(z_1)\overline{{\cal T}}_{n_e}(z_2)
\overline{{\cal T}}_{n_e}(z_3){\cal T}_{n_e}(z_4)\right\rangle_\mathbb{C}
=\exp\left[\frac{c}{8R}\left({\cal L}_{13}+{\cal L}_{24}-{\cal L}_{14}-{\cal L}_{23}\right)\right].
\end{equation}
We now utilize \cref{negativity,en,HenConjecture} and the Brown Henneaux formula $c=3R/(2G_N^{(3)})$ \cite{Brown:1986nw}, to express the holographic entanglement negativity for the mixed state configuration of the two disjoint intervals in proximity as\footnote{Note that the expression for the holographic entanglement negativity already contains the backreaction factor $\mathcal{X}_2=3/2$ \cite{Kudler-Flam:2018qjo} arising from the bulk cosmic brane appropriate to the \ads{3}/\cft{2} scenario.}
\begin{align}\label{en-conjecture}
{\cal E}&=\frac{3}{16G_N^{(3)}}
\left({\cal L}_{A_1\cup A_s}+{\cal L}_{A_2\cup A_s}-{\cal L}_{A_1\cup A_2\cup A_s}-{\cal L}_{A_s}\right)\\
&=\frac{3}{4}\left(S_{A_1\cup A_s}+S_{A_s\cup A_2}-S_{A_1\cup A_2\cup A_s}-S_{A_s}\right).
\end{align}
\begin{figure}
\centering
\includegraphics[scale=.40]{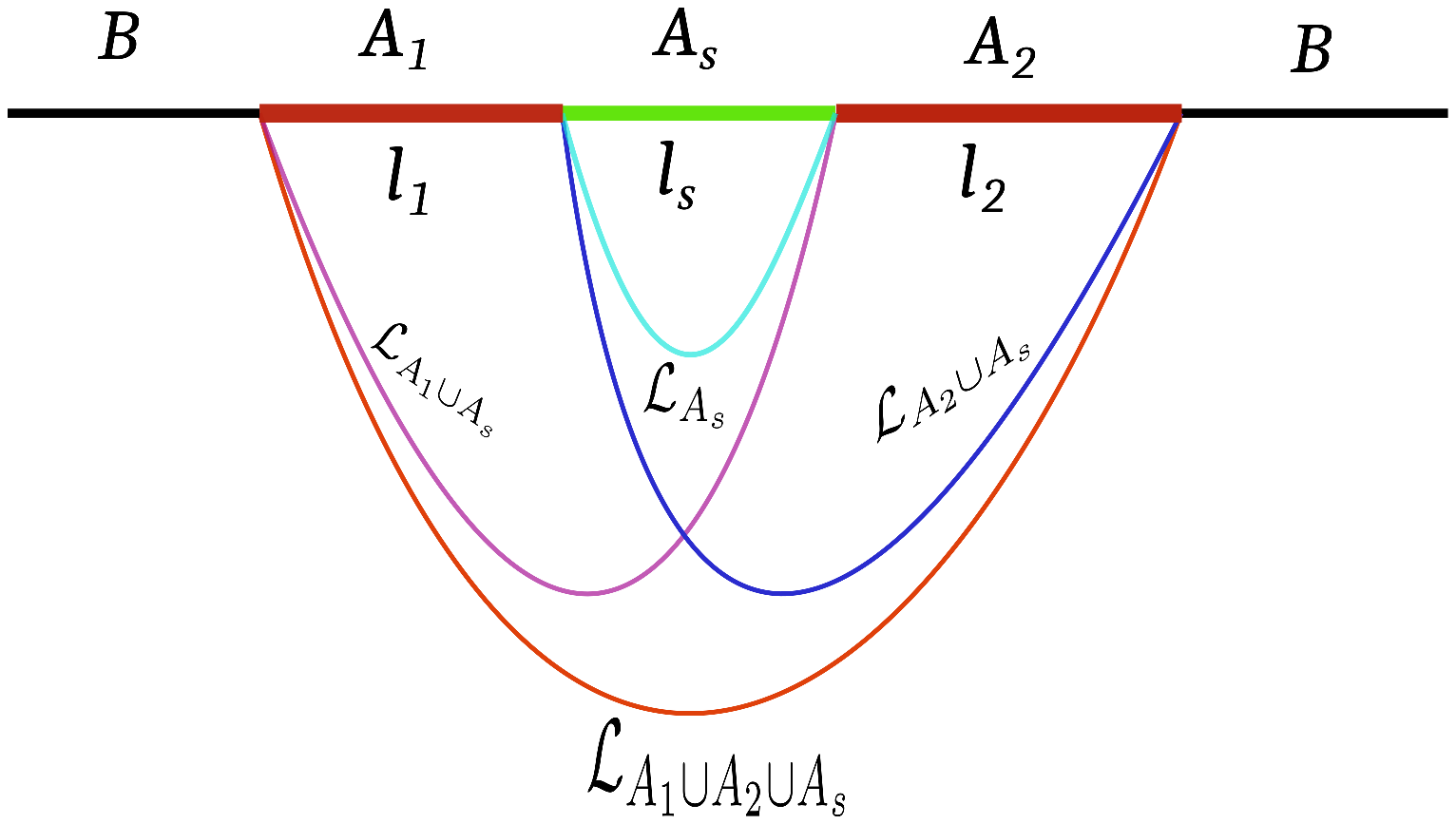}
\caption{Geodesics homologous to the intervals $A_1\cup A_s$, $A_s\cup A_2$, $A_1\cup A_s\cup A_2$ and $A_s$.}
\label{fig_hen_dj}
\end{figure}

The geodesics in \cref{en-conjecture} have been shown in \cref{fig_hen_dj}. Now the holographic entanglement negativity for the mixed state of disjoint intervals in the dual \cft{2} at zero temperature may be obtained using the holographic entanglement negativity conjecture described in \cref{en-conjecture}. The corresponding bulk configuration is described by a pure \ads{3} geometry which is expressed in the Poincar{\'e} coordinates as 
\begin{equation}\label{poin-metric}
ds^2=\left(\frac{r^2}{R^2}\right)\left(-dt^2+dx^2\right)+\left(\frac{r^2}{R^2}\right)^{-1}dr^2,
\end{equation}
where $R$ is the \ads{3} radius. The length ${\cal L}_\gamma$ of the bulk space like geodesic homologous to an interval $\gamma$ (of length $l_\gamma$) in these coordinates is written as \cite{Ryu:2006bv,Ryu:2006ef,Cadoni:2010ztg,Cadoni:2010kla}
\begin{equation}\label{length-pure}
{\cal L}_\gamma=2R\ln\left(l_\gamma/a\right),
\end{equation}
where $a$ is the UV cutoff for the \cft{2}. Now it is possible to utilize the expression in \cref{length-pure} to obtain the holographic entanglement negativity for the zero temperature mixed state in question from \cref{en-conjecture} in the following form \cite{Malvimat:2018txq}
\begin{equation}\label{en-vac}
{\cal E}=\frac{3R}{8G_N^{(3)}}\ln\left[\frac{(l_1+l_s)(l_2+l_s)}{l_s(l_1+l_2+l_s)}\right].
\end{equation}
On using the Brown Henneaux formula \cite{Brown:1986nw}, the above result exactly matches with the \cft{2} replica technique results for the large central charge limit as given in \cref{en_disj_vac}.

The corresponding holographic entanglement negativity for the finite temperature mixed state configuration of disjoint intervals (in proximity) in a \cft{2} may also be computed as above. For this case the dual bulk configuration is described by the Euclidean BTZ black string at a temperature $T$ \cite{Ryu:2006bv,Ryu:2006ef,Cadoni:2010ztg,Cadoni:2010kla} the metric for which is as follows
\begin{equation}\label{btz}
ds^2=\frac{\left(r^2-r_{h}^2\right)}{R^2}d\tau^2+\frac{R^2}{\left(r^2-r_{h}^2\right)}dr^2+\frac{r^2}{R^2}d\phi^2.
\end{equation}
In the above equation $\tau$ represents the Euclidean time where the $\phi$ coordinate is uncompactified and $r=r_h$ denotes the event horizon. The corresponding length ${\cal L}_{\gamma}$ of the bulk space like geodesic homologous to an interval $\gamma$ (of length $l_{\gamma}$) for this geometry may be expressed as \cite{Ryu:2006bv,Ryu:2006ef,Cadoni:2010ztg,Cadoni:2010kla} 
\begin{equation}\label{btz-geodesic}
{\cal L}_{\gamma}=2R\ln\left[\frac{\beta}{\pi a}\sinh\left(\frac{\pi l_{\gamma}}{\beta}\right)\right],
\end{equation}
where $a$ is the UV cutoff. Utilizing \cref{en-conjecture,btz-geodesic}, the holographic entanglement negativity for the finite temperature mixed state of disjoint intervals in proximity may be computed as
\cite{Malvimat:2018txq}
\begin{equation}\label{btz-en}
{\cal E}=\frac{3R}{8G_N^{(3)}}\ln
\left[\frac{\sinh\{\pi(l_1+l_s)/\beta\}\sinh\{\pi(l_2+l_s)/\beta\}}
{\sinh\{\pi l_s/\beta\}\sinh\{\pi(l_1+l_2+l_s)/\beta\}}\right].
\end{equation}
As earlier the holographic entanglement negativity in the above equation matches with the replica technique results in the large central charge limit described in \cref{cft-finite-temp} on utilizing the Brown Henneaux formula \cite{Brown:1986nw}.

\section{Holographic entanglement negativity for \ads{d+1}/\cft{d}}\label{sn_hen_hd}

Following the computation of the holographic entanglement negativity reviewed in the last section for mixed states of disjoint intervals in the \ads{3}/\cft{2} scenario \cite{Malvimat:2018txq,Malvimat:2018ood}, we proceed to propose a higher dimensional generalization in a generic \ads{d+1}/\cft{d} framework. The construction for the \ads{3}/\cft{2} scenario suggests that the corresponding higher dimensional construction would involve a similar algebraic sum of the areas of bulk RT surfaces\footnote {Note that for the corresponding \ads{3}/\cft{2} scenario these are geodesics homologous to appropriate intervals.} for the respective subsystems. The holographic entanglement negativity for such mixed state configurations in the \ads{d+1}/\cft{d} scenario may then be expressed as follows\footnote{\label{chi_d}Note that the right hand side of \cref{higher-d-conjecture} would be modified to incorporate the relevant backreaction effects. For spherical entangling surfaces, this entails replacing the numerical factor of $3/16$ by $\mathcal{X}_d/8$. See \cite{KumarBasak:2020ams} for details.}
\begin{equation}\label{higher-d-conjecture}
\mathcal{E}=\frac{3}{16G_N^{(d+1)}}(\mathcal{A}_{1s}+\mathcal{A}_{s2}-\mathcal{A}_{12s}-\mathcal{A}_s),
\end{equation}
where $\mathcal{A}_{ij}$ and $\mathcal{A}_{ijk}$ are the areas of the bulk RT surfaces for the subsystems $A_i\cup A_j$ and $A_i\cup A_j\cup A_k$ respectively with $i=1,2,s$, as depicted in \cref{fig_henhd_zero}. Using the RT prescription \cite{Ryu:2006bv}, the above expression for the holographic entanglement negativity may be expressed as follows%
\footnote{The right hand side of \cref{hen-hee-hd} would be likewise modified as detailed in \cref{chi_d}.}
\begin{equation}\label{hen-hee-hd}
\mathcal{E}=\frac{3}{4}(S_{A_1\cup A_s}+S_{A_s\cup A_2}-S_{A_1\cup A_2\cup A_s}-S_{A_s}) . 
\end{equation}
In the limit $A_s\to\emptyset$ (where $\emptyset$ is the null set), we recover the holographic entanglement negativity for the mixed state of adjacent subsystems as described in \cite{Jain:2017xsu}. This serves as a strong indication for the overall consistency of our proposal for the holographic entanglement negativity for such mixed state configurations in the \ads{d+1}/\cft{d} scenario.

As mentioned in the \nameref{sn_intro}, a related holographic construction for the entanglement negativity of bipartite states in \cft{d}s involving the EWCS backreacted by the bulk cosmic brane for the conical defect in the replicated geometry, has been proposed in \cite{Kudler-Flam:2018qjo,Kusuki:2019zsp} and further refined in \cite{KumarBasak:2020eia}. For spherical entangling surfaces the backreaction of the cosmic brane may be accounted for by an overall dimension dependent constant numerical factor obtained from the pure vacuum state in the \cft{d} dual to the bulk pure \ads{d+1} geometry \cite{Rangamani:2014ywa} although an explicit determination of this factor for arbitrary subsystem geometries is an extremely involved open issue. Considering these developments, we mention here that the higher dimensional proposal in \cite{Chaturvedi:2016rft,Jain:2017xsu,Jain:2018bai} and in this article, will also involve appropriate backreaction effects. However as mentioned above the determination of these effects for the subsystem geometries of long rectangular strips utilized in \cite{Chaturvedi:2016rft,Jain:2017xsu,Jain:2018bai} and the present article is a non trivial outstanding issue.\footnote{For more recent progress in these issues see \cite{KumarBasak:2020ams,Dong:2021clv,KumarBasak:2021lwm}.}

\section{Holographic entanglement negativity for \ads{d+1}/\cft{d} in vacuum}\label{sn_hen_hd_zero}

\begin{figure}
\centering
\includegraphics[scale=.40]{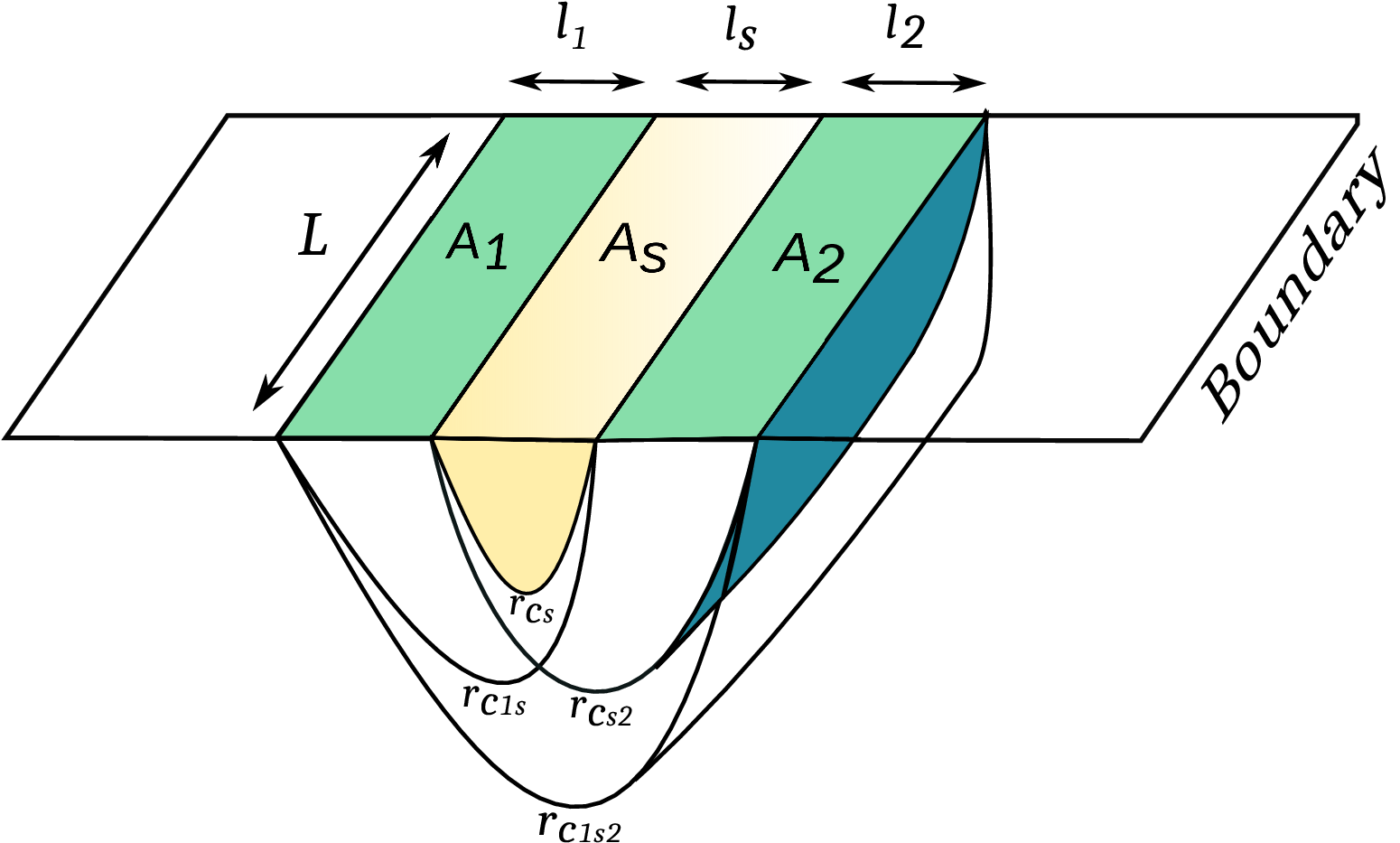}
\caption{RT surfaces for the subsystems $A_1\cup A_s$, $A_s\cup A_2$, $A_1\cup A_s\cup A_2$ and $A_s$ in the boundary \cft{d} at zero temperature.}
\label{fig_henhd_zero}
\end{figure}

In this section we employ our holographic conjecture described in the last section to compute the entanglement negativity for the zero temperature mixed state of two disjoint subsystems (in proximity) in a dual \cft{d}. The subsystems are described by $(d-1)$ dimensional spatial long rectangular strip geometries. In this case, the corresponding bulk geometry is the pure \ads{d+1} space time whose metric in Poincar{\'e} coordinates is given as
\begin{equation}
ds^2=\frac{1}{z^2}\Big(-dt^2+dz^2+\sum_{i=1}^{d-1}dx_i^2\Big).
\end{equation} 
The corresponding rectangular strip geometries are specified by the subsystems $A_1$, $A_2$ and $A_s$ as shown in \cref{fig_henhd_zero} with
\begin{equation}
x\equiv x^1=[-l_j/2,l_j/2],\quad x^i=[-L/2,L/2];\quad i=2,3,\dots,(d-1),\quad j=1,2,s;
\end{equation}
with
\begin{equation}\label{long}
L\gg l_1,l_2,l_s.
\end{equation}
Note that $x^1$ is known as the partitioning direction, and the other spatial directions are called transverse directions.
As mentioned in \cref{fn_long}, the conditions in \cref{long} guarantee that the subsystems $A_1$, $A_2$ and $A_s$ are long rectangular strips.

The proximity regime in higher dimensions is characterized by the condition that the separation between the two disjoint subsystems along the partitioning direction is much smaller than the lengths of the subsystems along that direction, given by
\begin{equation}\label{proximity}
l_s \ll l_1,l_2.
\end{equation}

We now briefly review the calculation for the area of a bulk RT surface such rectangular strip geometries described in \cite{Fischler:2012ca}. The area functional in this case may be expressed as follows
\begin{equation}\label{areaint}
\mathcal{A}=L^{d-2}\int_{-l/2}^{l/2}dx\frac{\sqrt{1+(\frac{dz}{dx})^2}}{z^{d-1}}.
\end{equation}
Upon extremization of the area functional in \cref{areaint} we arrive at the following differential equation
\begin{equation}\label{diffz}
\frac{dz}{dx}=\frac{\sqrt{z_*^{2(d-1)}-z^{2(d-1)}}}{z^{2(d-1)}}.
\end{equation}
Here, $z=z_*$ is the turning point of the static minimal surface. Utilizing \cref{areaint,diffz}, the area functional may be expressed as
\begin{equation}\label{totA}
\mathcal{A}=\mathcal{A}_{div}+\mathcal{A}_{finite},
\end{equation}
where
\begin{equation}
\mathcal{A}_{div}=\frac{2}{d-2}\left(\frac{L}{a}\right)^{d-2},
\end{equation}
\begin{equation}
\mathcal{A}_{finite}=\frac{2\sqrt{\pi}\Gamma\Big(\frac{-d+2}{2d-2}\Big) }{\Gamma\Big(\frac{1}{2d-2}\Big)}\Big(\frac{L}{z_*}\Big)^{d-2}
=S_0\left(\frac{L}{l}\right)^{d-2},
\end{equation}
with $a$ being the UV cutoff and the constant $S_0$ is given as
\begin{equation}\label{S0}
S_0=-\frac{2^{d-1}\pi^{\frac{(d-1)}{2}}}{d-2}\Bigg(\frac{\Gamma\Big(\frac{d}{2d-2}\Big)}{\Gamma\Big(\frac{1}{2d-2}\Big)}\Bigg)^{d-1}.
\end{equation}

We now utilize \cref{higher-d-conjecture} to compute the holographic entanglement negativity for the zero temperature mixed state of the disjoint long rectangular strip geometries in proximity as
\begin{equation}\label{neg}
\mathcal{E}=\frac{3S_0}{16G_N^{(d+1)}}
\Bigg[\Big(\frac{L}{l_1+l_s}\Big)^{d-2}+\Big(\frac{L}{l_2+l_s}\Big)^{d-2}-\Big(\frac{L}{l_1+l_2+l_s}\Big)^{d-2}-\Big(\frac{L}{l_s}\Big)^{d-2}\Bigg].
\end{equation} 
Note that the above result is cutoff independent as expected, unlike the case for the mixed state of adjacent subsystems as described in \cite{Jain:2017xsu}.

In the limit $l_s\to a$ in \cref{neg}, where the subsystems become adjacent, we arrive at the following expression for the holographic entanglement negativity
\begin{equation}
\mathcal{E}_{Adjacent}=\frac{3}{16 G_N^{(d+1)}}\Bigg[\frac{2}{d-2}\Big(\frac{L}{a}\Big)^{d-2}+S_0\bigg\{\Big(\frac{L}{l_1}\Big)^{d-2}+\Big(\frac{L}{l_2}\Big)^{d-2}-\Big(\frac{L}{l_1+l_2}\Big)^{d-2}\bigg\}\Bigg].
\end{equation} 
Note that in arriving at the above expression we have added and subtracted the divergent part of the entanglement negativity from the right hand side of \cref{neg} and neglected sub leading terms in the limit $l\gg a$. Interestingly the above result exactly matches with the corresponding holographic entanglement negativity for adjacent subsystems as described in \cite{Jain:2017xsu} which serves as a strong consistency check for our construction.

\section{Holographic entanglement negativity for \ads{d+1}/\cft{d} at finite temperatures}\label{sn_hen_hd_finite}

Having computed the holographic entanglement negativity for the zero temperature mixed state of two disjoint subsystems
in proximity we turn our attention in this section to the corresponding finite temperature case where the dual bulk geometry is described by the \ads{d+1}-Schwarzschild black hole. The metric for the \ads{d+1}-Schwarzschild black hole with the AdS radius $R=1$ is given as
\begin{equation}
ds^2= -r^2\left(1-\frac{r_h^{ d}}{r^{d}}\right)dt^2+\frac{dr^2}{r^2\Big(1-\frac{r_h^{d}}{r^{d}}\Big)}+r^2d\vec{x}^2,
\end{equation}
where $\vec{x}\equiv(x,x^i)$ are the coordinates on the boundary and the horizon radius $r_h$ is related to Hawking temperature as $r_h=4\pi T/d$.

We begin by briefly reviewing the computation of the area of the bulk RT surface for a subsystem of long rectangular strip geometry on the boundary in this case, as described in \cite{Fischler:2012ca}. The area functional for a single long rectangular strip in this case may be expressed as
\begin{equation}\label{5areaint}
{\cal A}=L^{d-2}\int dr\;r^{d-2}\sqrt{r^2x'^2+\frac{1}{r^2\left(1-\frac{r_h^d}{r^d}\right)}}.
\end{equation}
Extremizing the above area functional we obtain
\begin{equation}\label{l/2}
\frac{l}{2}=\frac{1}{r_c}\int_0^1\frac{u^{d-1} du}{\sqrt{(1-u^{2d-2})}}
\left(1-\frac{r_h^d}{r_{c}^d}u^d\right)^{-\frac{1}{2}},\quad u=\frac{r_c}{r},
\end{equation} 
where $r_c$ is the turning point as earlier. The area functional in \cref{5areaint} may now be re-expressed in terms of the variable $u$ as
\begin{equation}\label{ar}
{\cal A}=2L^{d-2}r_c^{d-2}\int_0^1\frac{du}{u^{d-1}\sqrt{(1-u^{2d-2})}}
\left(1-\frac{r_h^d}{r_{c}^d}u^d\right)^{-\frac{1}{2}}.
\end{equation}

The above integral diverges at the lower limit and as earlier we express the area functional in the following form
\begin{equation}\label{fullarea}
{\cal A}=\mathcal{A}_{div}+\mathcal{A}_{finite},
\end{equation}
where $\mathcal{A}_{div}$ is the temperature independent divergent part and $\mathcal{A}_{finite}$ is the temperature dependent finite part. The divergent and the finite parts may be written as
\begin{equation}\label{A_div}
\mathcal{A}_{div}=\frac{2}{d-2}\Big(\frac{L}{a}\Big)^{d-2},
\end{equation}
\begin{equation}\label{A_fin}
\mathcal{A}_{finite}= L^{d-2}r_c^{d-2}\Bigg[\mathcal{P}+\sum_{n=1}^{\infty}\mathcal{Q}_n\Big(\frac{r_h}{r_c}\Big)^{nd} \Bigg],
\end{equation}
where the series on the right hand side of \cref{A_fin} always converges for $r_c>r_h$ \cite{Hubeny:2012ry}, and the expressions for $\mathcal{P}$ and $\mathcal{Q}_n$ are given by
\begin{equation}
\mathcal{P}=\frac{\sqrt{\pi}\Gamma\Big(-\frac{d-2}{2(d-1)}\Big)}{(d-1)\Gamma\Big(\frac{1}{2(d-1)}\Big)},
\quad\mathcal{Q}_n=\Big(\frac{1}{(d-1)}\Big)\frac{\Gamma\Big(n+\frac{1}{2}\Big)\Gamma\Big(\frac{d(n-1)+2}{2(d-1)}\Big)}{\Gamma\big(1+n\big)\Gamma\Big(\frac{dn+1}{2(d-1)}\Big)}.
\end{equation}
Finally utilizing our conjecture described in \cref{higher-d-conjecture} the holographic entanglement negativity for the mixed state configuration of two disjoint subsystems of long rectangular strip geometries in question may be expressed as
\begin{align}
\mathcal{E}=\frac{3L^{d-2}}{16G_N^{(d+1)}}\Bigg[
&r_{c_{1s}}^{d-2}\Bigg\{\mathcal{P}+\sum_{n=1}^{\infty} \mathcal{Q}_n
\Big(\frac{r_h}{r_{c_{1s}}}\Big)^{nd} \Bigg\} 
+r_{c_{s2}}^{d-2}\Bigg\{\mathcal{P}+
\sum_{n=1}^{\infty} \mathcal{Q}_n
\Big(\frac{r_h}{r_{c_{s2}}}\Big)^{nd} \Bigg\}\nonumber\\
&-r_{c_{12s}}^{d-2}\Bigg\{\mathcal{P}+
\sum_{n=1}^{\infty} \mathcal{Q}_n
\Big(\frac{r_h}{r_{c_{12s}}}\Big)^{nd} \Bigg\} 
-r_{c_s}^{d-2}\Bigg\{\mathcal{P}+
\sum_{n=1}^{\infty} \mathcal{Q}_n
\Big(\frac{r_h}{r_{c_s}}\Big)^{nd} \Bigg\} \Bigg].
\end{align}
Here $r_{c_{1s}}$, $r_{c_{s2}}$, $r_{c_{12s}}$, $r_{c_s}$ describe the turning points of the static minimal surfaces in the bulk as depicted in \cref{fig_henhd_zero}. Note that the quantities $r_{c_{ij}}$, $r_{c_{ijk}}$ and $r_{c_{i}}$ in the above equation are required to be expressed in terms of the subsystem lengths $l_{ij}$, $l_{ijk}$, $l_{i}$ and $r_h$ using the integral described in \cref{l/2}, where $i,j=1,2,s$. In the next section we compute this integral in a perturbative approximation for low and high temperatures to extract the leading contribution to the holographic entanglement negativity.

\subsection{Holographic entanglement negativity in the low temperature limit}

The low temperature limit for the integral in \cref{l/2} is described by the regime $Tl\ll 1$ (or $r_hl\ll 1$) as the turning point for the static minimal surface remains far away from the horizon at $r_h$. The quantity $r_c$ describing the turning point for the static minimal surface in the bulk described in \cref{l/2} may now be evaluated perturbatively as a series expansion in $r_hl$. The finite part of the area in this case may be expressed as \cite{Fischler:2012ca}
\begin{equation}\label{Afinlow}
\mathcal{A}_{finite}=S_0\Big(\frac{L}{l}\Big)^{d-2}\bigg[1+S_1(r_hl)^d+O[(r_hl)^{2d}]\bigg].
\end{equation}
Here $S_0$ is the same constant as given in \cref{S0} and $S_1$ is another constant given by
\begin{equation}\label{S1}
S_1=\frac{\Gamma\big(\frac{1}{2(d-1)}\big)^{d+1}}{2^{d+1}\pi^{\frac{d}{2}}\Gamma\big(\frac{d}{2(d-1)}\big)^d\Gamma\big(\frac{d+1}{2(d-1)}\big)}\bigg(\frac{\Gamma\big(\frac{1}{d-1}\big)}{\Gamma\big(-\frac{d-2}{2(d-1)}\big)}+\frac{2^\frac{1}{d-1}(d-2)\Gamma\big(1+\frac{1}{2(d-1)}\big)}{\sqrt{\pi}(d+1)}\bigg).
\end{equation}

\begin{figure}
\centering
\includegraphics[scale=.40]{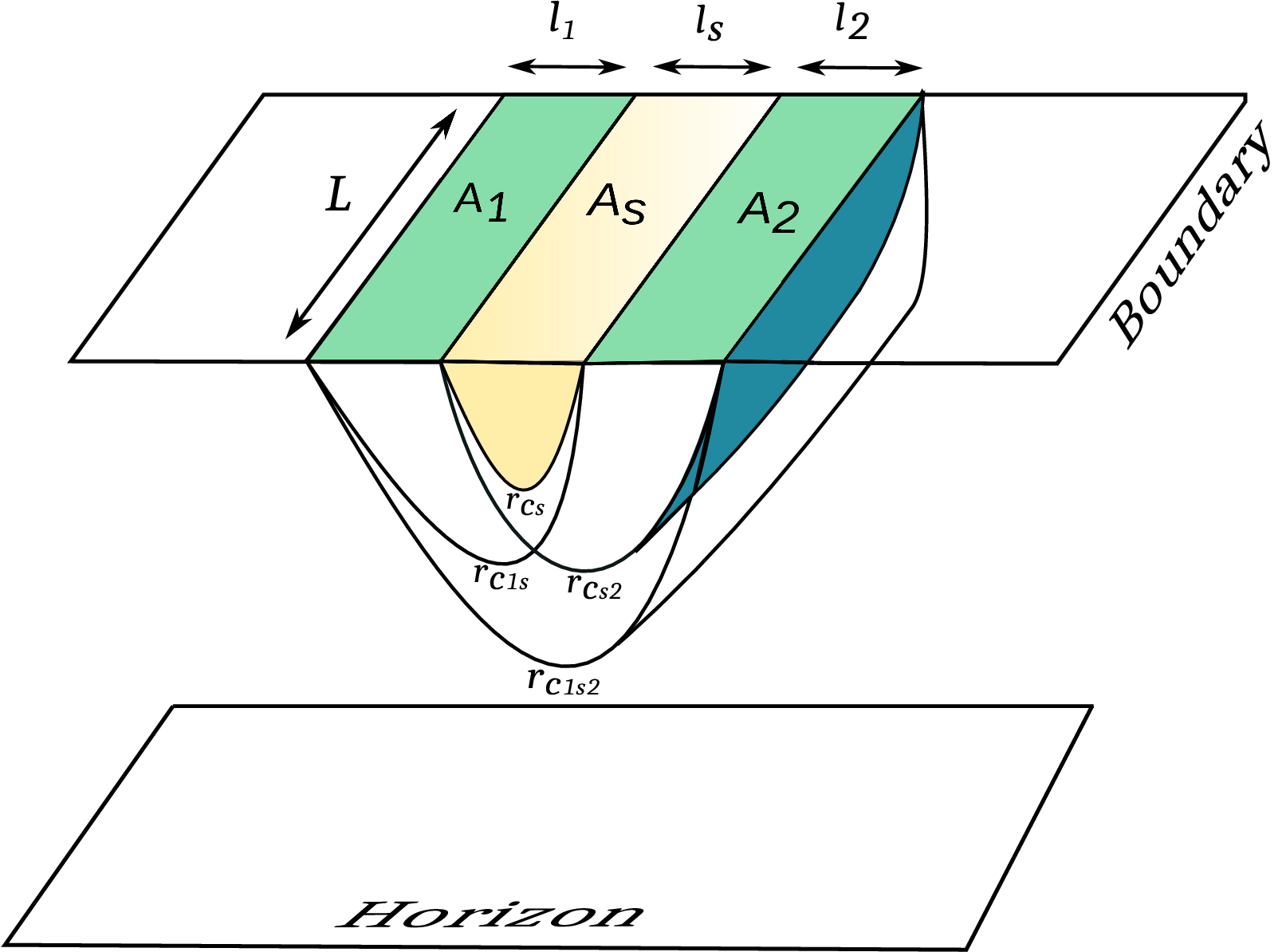}
\caption{RT surfaces for the subsystems $A_1\cup A_s$, $A_s\cup A_2$, $A_1\cup A_s\cup A_2$ and $A_s$ in the boundary \cft{d} at low temperature.}
\label{fig_henhd_low}
\end{figure}

The holographic entanglement negativity at low temperatures for the mixed state of two disjoint subsystems with long rectangular strip geometries as described in \cref{fig_henhd_low} may now be perturbatively computed utilizing our conjecture in \cref{higher-d-conjecture}, and \cref{fullarea,Afinlow,S1} as follows
\begin{align}
\mathcal{E}=\frac{3}{16 G_N^{(d+1)}}\Bigg[&S_0\Bigg\{\Bigg(\frac{L}{l_1+l_s}\Bigg)^{d-2}+\Bigg(\frac{L}{l_2+l_s}  \Bigg)^{d-2}-\Bigg(\frac{L}{l_1+l_2+l_s}\Bigg)^{d-2}-\Bigg(\frac{L}{l_s}\Bigg)^{d-2}\Bigg\}\nonumber\\  
&+S_0S_1L^{d-2}\Big(\frac{4\pi T}{d}\Big)^d\Big\{(l_1+l_s)^2+(l_2+l_s)^2-(l_1+l_2+l_s)^2-l_s^2\Big\}\nonumber\\
&+S_0 \Bigg\{\Big(\frac{L}{l_1+l_s}\Big)^{d-2}\mathcal{O}(T(l_1+l_s))^{2d} +\Big(\frac{L}{l_2+l_s}\Big)^{d-2}\mathcal{O}(T(l_2+l_s))^{2d}\nonumber\\ &-\Big(\frac{L}{l_1+l_2+l_s}\Big)^{d-2}\mathcal{O}(T(l_1+l_2+l_s))^{2d}\Bigg\}-S_0\Big(\frac{L}{l_s}\Big)^{d-2}\mathcal{O}(Tl_s)^{2d}\Bigg].\label{entlow}
\end{align}
Note that this result is also cutoff independent in contrast with the case for the mixed state configuration of adjacent intervals in \cite{Jain:2017xsu}. The first term on the right hand side of the above equation arises from the contribution of the \ads{d+1} vacuum described in \cref{neg} and is temperature independent. The remaining terms are the finite temperature corrections to the holographic entanglement negativity at low temperatures which is similar to the case of the mixed state of adjacent intervals as reported in \cite{Jain:2017xsu}.

Using similar arguments as described in \cref{sn_hen_hd_zero} we may obtain the corresponding holographic entanglement negativity for the mixed state of two adjacent subsystems at low temperature through the adjacent limit $l_s\to a$ in \cref{entlow} as
\begin{align}
\mathcal{E}_{Adjacent}=\frac{3}{16 G_N^{(d+1)}}
\Bigg[&\frac{2}{d-2}\Big(\frac{L}{a}\Big)^{d-2}+S_0\Bigg\{\Big(\frac{L}{l_1}\Big)^{d-2}+\Big(\frac{L}{l_2}\Big)^{d-2}-\Big(\frac{L}{l_1+l_2}\Big)^{d-2}\Bigg\}\nonumber\\
&-kl_1 l_2L^{d-2}T^d+S_0\Bigg\{\Big(\frac{L}{l_1}\Big){\cal O}\big(Tl_1\big)^{2d}+\Big(\frac{L}{l_2}\Big){\cal O}\big(Tl_2\big)^{2d}\Bigg\}\Bigg],
\end{align}
where $k=2(4\pi/d)^dS_0S_1$ is a constant. As earlier for the zero temperature mixed state dual to the bulk pure \ads{d+1} geometry, the above result matches with the corresponding holographic entanglement negativity for adjacent intervals \cite{Jain:2017xsu} at low temperatures further validating our construction.

\subsection{Holographic entanglement negativity in the high temperature limit}

\begin{figure}
\centering
\includegraphics[scale=.40]{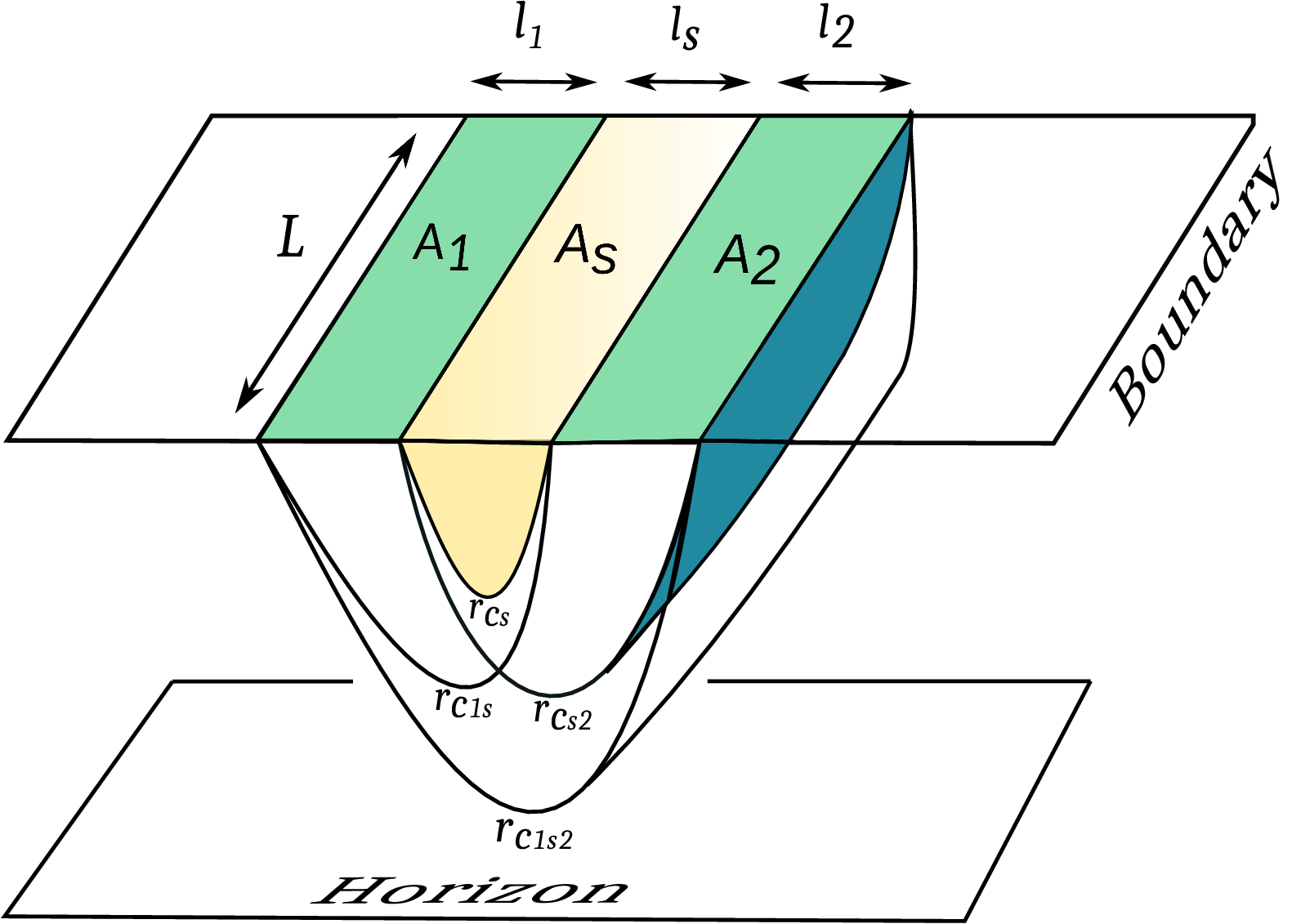}
\caption{RT surfaces for the subsystems $A_1\cup A_s$, $A_s\cup A_2$, $A_1\cup A_s\cup A_2$ and $A_s$ in the boundary \cft{d} at high temperature.}
\label{fig_henhd_high}
\end{figure}

We now proceed to calculate the entanglement negativity for the mixed state of disjoint subsystems as depicted in \cref{fig_henhd_high} at high temperature limit $Tl\gg 1$ ($r_hl\gg 1$). In the above limit the bulk RT surface for a subsystem in the dual \cft{d} approaches the black hole horizon, hence the turning point radius $r_c$ is large and $r_c \approx r_h$.

As described in \cite{Fischler:2012ca} the integral in \cref{l/2} can be expanded in terms of 
$\epsilon= r_c/r_h-1$ and solving it up to the leading order leads to the following expression 
\begin{equation}\label{epsilon}
\epsilon=C_1\exp(-\sqrt{\frac{d(d-1)}{2}}\;lr_h).
\end{equation}
Here $C_1$ is a constant which is given by
\begin{align}
C_1=\frac{1}{d}\ \exp\Bigg[&\sqrt{\frac{d(d-1)}{2}}\Bigg\{\frac{2\sqrt{\pi}\Gamma\Big(\frac{d}{2(d-1)}\Big)}{\Gamma\Big(\frac{1}{2(d-1)}\Big)}\nonumber\\&+2\sum_{n=1}^{\infty} \Bigg(\frac{1}{1+nd}\frac{\Gamma\Big(n+\frac{1}{2}\Big)\Gamma\Big(\frac{d(n+1)}{2(d-1)}\Big)}{\Gamma\Big(1+n\Big)\Gamma\Big(\frac{dn+1}{2(d-1)}\Big)}-\frac{1}{\sqrt{2d(d-1)}~n}\Bigg)\Bigg\} \Bigg].\label{c1}
\end{align}
The area of a bulk static minimal surface homologous to the subsystem in question at a high temperature may then be obtained in terms of $r_h=4\pi T/d$ by writing \cref{A_fin} as an expansion of $\epsilon$ and using \cref{fullarea,A_div,epsilon} as follows
\begin{align}
\mathcal{A}=&\frac{2}{d-2}\Big(\frac{L}{a}\Big)^{d-2}+\Big(\frac{4\pi}{d}\Big)^{d-1}\Bigg[V~T^{d-1}+\frac{C_2 ~d}{8\pi}A'~T^{d-2}\nonumber\\
&-\frac{C_1}{8\pi}\sqrt{2d(d-1)}~A'~T^{d-2} \exp\Big\{-\sqrt{(d-1)/2d}~4 \pi Tl\Big\}+\cdots\Bigg],\label{Ahigh}
\end{align}
where $V=l\,L^{d-2}$ is the volume of the subsystem and $A'=2L^{d-2}$ is the area of a single long rectangular strip. Here $C_2$ is another constant which is given as
\begin{equation}
C_2=2\Bigg[-\frac{\sqrt{\pi}(d-1)\Gamma\Big(\frac{d}{2(d-1)}\Big)}{(d-2)\Gamma\Big(\frac{1}{2(d-1)}\Big)}
+\sum_{n=1}^{\infty}\frac{1}{1+nd}\Big(\frac{d-1}{d(n-1)+2}\Big)\frac{\Gamma\Big(n+1/2\Big)
\Gamma\Big(\frac{d(n+1)}{2d-2}\Big)}{\Gamma\big(n+1\big)\Gamma\Big(\frac{dn+1}{2d-2}\Big)}\Bigg].
\end{equation}

The holographic entanglement negativity at high temperatures for the mixed state of disjoint subsystems of long rectangular strip geometries may now be obtained from our conjecture by utilizing \cref{higher-d-conjecture,Ahigh} in the following form
\begin{align}
\mathcal{E}=&\frac{3}{16G_N^{(d+1)}}\left(\frac{4\pi}{d}\right)^{d-1}\frac{C_1}{4\pi}\sqrt{2d(d-1)}\,A\,T^{d-2}\Bigg[-\exp\Big\{ -\sqrt{\frac{d-1}{2d}}4\pi T(l_1+l_s)\Big\}\nonumber\\
&-\exp\Big\{-\sqrt{\frac{d-1}{2d}}4\pi T(l_2+l_s)\Big\}
+\exp\Big\{-\sqrt{\frac{d-1}{2d}}4\pi T(l_1+l_2+l_s)\Big\}\nonumber\\
&+\exp\Big\{-\sqrt{\frac{d-1}{2d}}4\pi Tl_s\Big\}+\cdots\Bigg].\label{Ehigh}
\end{align}
Here $A=L^{d-2}$ is the area of the entangling surface between the two adjacent long rectangular strips (in proximity)
in the dual \cft{d} and the ellipses represent the higher order correction terms. As earlier this result is also cutoff independent. Note that in the high temperature limit the volume dependent thermal terms in \cref{Ahigh} cancel between the two disjoint subsystems leading to an expression for the holographic entanglement negativity that is proportional to the transverse area of the subsystems at the leading order in the perturbative expansion. This is expected from a quantum information perspective as the entanglement negativity characterizes an upper bound to the distillable entanglement for the mixed state under consideration and should not involve volume dependent thermal contributions. Note that the subtraction for the thermal part at the leading order in this case is more subtle than for the mixed state configuration of a single interval at a finite temperature described in \cite{Chaturvedi:2016rft}.

Following a similar procedure as described earlier for the low temperature regime, it is possible to obtain the holographic entanglement negativity in the limit when the two subsystems are adjacent with $l_s\to a$ as

\begin{align}
{\cal E}_{Adjacent}=&\frac{3}{16 G_N^{(d+1)}}\frac{2}{(d-2)}\Big(\frac{A}{a^{d-2}}\Big)+\frac{3}{16 G_N^{(d+1)}}\Big(\frac{4\pi}{d}\Big)^{d-1}\Bigg[\frac{C_2d}{4\pi}AT^{d-2}\nonumber\\
&-\frac{C_1}{4\pi}\sqrt{2d(d-1)}AT^{d-2}\Bigg\{\exp\Big(-\sqrt{(d-1)/2d}~4 \pi Tl_1\Big)\nonumber\\
&+\exp\Big(-\sqrt{(d-1)/2d}\,4 \pi Tl_2\Big)
-\exp\Big(-\sqrt{(d-1)/2d}\,4 \pi T(l_1+l_2)\Big)\Bigg\}+\cdots\Bigg].\label{Eadjhigh}
\end{align}
Once again this matches exactly with the adjacent interval results described in \cite{Jain:2017xsu} and constitutes a consistency check for our construction.

\section{Summary and conclusions}\label{sn_sum}

To summarize, we have advanced a construction for the holographic entanglement negativity of bipartite mixed states of disjoint subsystems in proximity, in \cft{d}s dual to bulk \ads{d+1} geometries. Our proposal arises from the corresponding \ads{3}/\cft{2} scenario for such mixed state configurations and involves an algebraic sum of the areas of bulk RT surfaces for certain combinations of subsystems relevant to the mixed state in the dual \cft{d}. As a consistency check we have obtained the holographic entanglement negativity for mixed states described by disjoint subsystems (in proximity) with long rectangular strip geometries in \cft{d}s dual to bulk pure \ads{d+1} and \ads{d+1}-Schwarzschild black hole geometries respectively at leading orders in a perturbation theory. In the latter case the area integrals for the RT surfaces were computed perturbatively for both the low and high temperature regimes.

It is observed from our results that at low temperatures the dominant contribution to the holographic entanglement negativity arises from the pure \ads{d+1} vacuum with sub leading finite temperature corrections. In the high temperature limit at leading orders we demonstrate a subtle cancellation between the volume dependent thermal contributions which renders the holographic entanglement negativity to be dependent on the area of the entangling surfaces. The elimination of the thermal terms conform to the characterization of entanglement negativity as an upper bound to the distillable entanglement in quantum information theory. Furthermore the results are cutoff independent as expected from the analysis of the corresponding lower dimensional \ads{3}/\cft{2} scenario. Interestingly we have exactly reproduced the results described earlier in the literature for mixed states of adjacent subsystems in a dual \cft{d} through an appropriate adjacent limit, which provides a further consistency check for our construction. As mentioned earlier a proof of our holographic construction valid for spherical entangling surfaces may be inferred from a recent communication involving replica symmetry breaking saddles for the bulk gravitational path integral. However such a proof for generic subsystem geometries in particular for the long rectangular strip geometries described in this article is still a non trivial but fascinating open issue which needs further investigation.

In this context as mentioned in the \nameref{sn_intro}, an alternative holographic entanglement negativity conjecture involving the entanglement wedge cross section (EWCS), backreacted by the cosmic brane for the conical defect in the replicated bulk geometry in a gravitational path integral, has been proposed in the literature.  For spherical entangling surfaces this backreaction results in a dimension dependent overall numerical constant for the holographic entanglement negativity of the dual \cft{d} in the replica limit. Applied to the \ads{3}/\cft{2} scenario this proposal reproduced the corresponding entanglement negativity for bipartite states in dual \cft{2}s in the large central charge limit modulo certain constants related to the holographic Markov gap described recently in the literature. It also reproduced the earlier holographic results for the \ads{3}/\cft{2} scenario up to the constants mentioned above, from the proposal based on the algebraic sums of the areas of the RT surfaces (lengths of geodesics in this case) for combinations of intervals relevant to the bipartite state under consideration. This suggests the equivalence of the two holographic constructions modulo the constants arising from the Markov gap as mentioned earlier. Note however that a higher dimensional extension of the alternative proposal involving the bulk EWCS is extremely non trivial and currently an open issue.

As mentioned in this work keeping the above developments in our perspective, we have stated that our higher dimensional results will also involve such relevant backreaction effects as mentioned above. The explicit determination of these effects for the long rectangular strip geometries for the subsystems considered in this article is a non trivial open issue which requires further investigation in the context of a generic higher dimensional \ads{d+1}/\cft{d} scenario. We hope to return to these fascinating open issues mentioned above in the near future which is expected to lead to interesting insights and the development of a holographic construction for the entanglement negativity for arbitrary subsystem geometries in higher dimensional \cft{d}s dual to bulk \ads{d+1} spacetimes and its proof from a bulk gravitational path integral perspective. This will have substantial significance for phenomena which involve mixed state entanglement in quantum field theories, quantum gravity and black holes.

\section*{Acknowledgments}

The work of JKB is supported by the National Science and Technology Council of Taiwan with the grant 112-2636-M-110-006. The work of HP is supported by the NCTS, Taiwan. The work of GS is partially supported by the Dr.\@ Jag Mohan Garg Chair Professor position at the Indian Institute of Technology, Kanpur.

\appendix

\section{A plausible entanglement temperature for holographic entanglement negativity}\label{app_ent_temp_hen}

In this appendix we briefly discuss a plausible entanglement temperature for the holographic entanglement negativity (HEN), along the lines of \cite{Bhattacharya:2012mi}, where the authors demonstrated that for a small subsystem, the holographic entanglement entropy (HEE) follows a ``first law'' like relation for excited states of the system, and the corresponding effective temperature is inversely proportional to the size of the subsystem under consideration. To this end we consider an excited state in a dual CFT$_d$ with subsystems $A$ and $B$ ($B=A^c$). The HEE/HEN of $A$ for this excited state is given by $S_A/\mathcal{E}_A$, and $S_A^{(0)}/\mathcal{E}_A^{(0)}$ denotes the HEE/HEN for the pure AdS$_{d+1}$ (which corresponds to the ground state of the dual CFT). The difference in HEE/HEN is given by $\Delta S_A\equiv S_A-S_A^{(0)}$, $\Delta\mathcal{E}_A\equiv \mathcal{E}_A-\mathcal{E}_A^{(0)}$, and the corresponding change in energy is indicated by $\Delta E_A$. It could then be shown that \cite{Bhattacharya:2012mi}
\begin{equation}\label{first_law_HEE}
T_{ent}\cdot \Delta S_A=\Delta E_A,
\end{equation}
where the effective temperature $T_{ent}$ is known as the entanglement temperature. We now attempt to find whether a similar thermodynamic relation exists for HEN, as follows
\begin{equation}\label{first_law_HEN}
T_{hen}\cdot \Delta\mathcal{E}_A=\Delta E_A.
\end{equation}
In what follows we derive an expression for $T_{hen}$ (the effective temperature for HEN) for a single subsystem with rectangular strip geometry in a thermal CFT$_d$ dual to an AdS$_{d+1}$-Schwarzschild black hole, in the low temperature regime. The HEN for a single subsystem A of length $l$ (along the partitioning direction) and transverse length $L$ for the configuration described above is given at the leading order by \cite{Chaturvedi:2016rft}
\begin{equation}\label{HENSG_FT}
\mathcal{E}_A=\frac{3}{8G_N^{(d+1)}}
\left[\frac{2}{d-2}\left(\frac{L}{a}\right)^{d-2}
+\left(\frac{L}{l}\right)^{d-2}\left\{S_0+S_0S_1\left(\frac{4\pi Tl}{d}\right)^d
-\left(\frac{4\pi Tl}{d}\right)^{d-1}\right\}\right].
\end{equation}
The ground state negativity may be obtained by taking $T\to 0$ in \cref{HENSG_FT} as follows
\begin{equation}\label{HENSG_ZT}
\mathcal{E}_A^{(0)}=\frac{3}{8G_N^{(d+1)}}
\left[\frac{2}{d-2}\left(\frac{L}{a}\right)^{d-2}
+S_0\left(\frac{L}{l}\right)^{d-2}\right].
\end{equation}
The change in negativity to the leading order is then computed as ($Tl\ll 1$)
\begin{equation}\label{HENSG_Del}
\Delta\mathcal{E}_A=\frac{3}{8G_N^{(d+1)}}\left(\frac{L}{l}\right)^{d-2}
\left[S_0S_1\left(\frac{4\pi Tl}{d}\right)^d
-\left(\frac{4\pi Tl}{d}\right)^{d-1}\right]
\approx -\frac{3}{8G_N^{(d+1)}}\left(\frac{L}{l}\right)^{d-2}
\left(\frac{4\pi Tl}{d}\right)^{d-1}.
\end{equation}
The corresponding change in energy is given by \cite{Bhattacharya:2012mi}
\begin{equation}\label{SG_DelE}
\Delta E_A=\frac{(d-1)T}{4dG_N^{(d+1)}}\left(\frac{L}{l}\right)^{d-2}
\left(\frac{4\pi Tl}{d}\right)^{d-1}.
\end{equation}
Substituting \cref{HENSG_Del,SG_DelE} into \cref{first_law_HEN} we arrive at
\begin{equation}\label{HEN_Temp_SG}
T_{hen}=-\frac{2(d-1)}{3d}T.
\end{equation}

Note that, unlike the HEE scenario, the effective temperature for HEN in \cref{HEN_Temp_SG} does not exhibit any scaling behavior with subsystem size. This is due to the presence of the (subtracted) volume dependent thermal entropy term for the HEN of a single subsystem. Interestingly $T_{hen}$ is proportional to the (real) temperature, and the proportionality factor depends only on dimensionality and shape of the subsystem (similar to the HEE case). Also note that $T_{hen}$ is negative, which is consistent with the fact that negativity decreases with increase in temperature (energy).

\bibliographystyle{JHEP}
\bibliography{references}
\end{document}